\documentclass[submitting]{nst}

\usepackage{subfigure,dcolumn}
\usepackage{epstopdf}
\usepackage{mhchem}
\usepackage{upgreek}
\usepackage{soul}

\begin{document}

\title{ CSHINE for studies of HBT correlation in Heavy Ion Reactions}\thanks{This work was supported by the National Natural Science Foundation of China (Nos. 11875174 and 11961131010). }

\author{Yi-Jie Wang}
\email[Corresponding author, ]{yj-wang15@mails.tsinghua.edu.cn}
\affiliation{Department of Physics, Tsinghua University, Beijing 100084, China}

\author{Fen-Hai Guan}
\affiliation{Department of Physics, Tsinghua University, Beijing 100084, China}

\author{Xin-Yue Diao}
\affiliation{Department of Physics, Tsinghua University, Beijing 100084, China}

\author{Qiang-Hua Wu}
\affiliation{Department of Physics, Tsinghua University, Beijing 100084, China}

\author{Xiang-Lun Wei}
\affiliation{Institute of Modern Physics, Lanzhou 730000, China}

\author{He-Run Yang}
\affiliation{Institute of Modern Physics, Lanzhou 730000, China}

\author{Peng Ma}
\affiliation{Institute of Modern Physics, Lanzhou 730000, China}

\author{Zhi Qin}
\affiliation{Department of Physics, Tsinghua University, Beijing 100084, China}

\author{Yu-Hao Qin}
\affiliation{Department of Physics, Tsinghua University, Beijing 100084, China}

\author{Dong Guo}
\affiliation{Department of Physics, Tsinghua University, Beijing 100084, China}

\author{Rong-Jiang Hu}
\affiliation{Institute of Modern Physics, Lanzhou 730000, China}

\author{Li-Min Duan}
\affiliation{Institute of Modern Physics, Lanzhou 730000, China}

\author{Zhi-Gang Xiao}
\affiliation{Department of Physics, Tsinghua University, Beijing 100084, China}

\begin{abstract}
 The Compact Spectrometer for Heavy Ion Experiment (CSHINE) is under construction for the study of isospin chronology via the Hanbury Brown$-$Twiss (HBT) particle correlation function and the nuclear equation of state of asymmetrical nuclear matter. The CSHINE consists of silicon strip detector (SSD) telescopes and large-area parallel plate avalanche counters, which measure the light charged particles and fission fragments, respectively. In phase I, two SSD telescopes were used to observe 30 MeV/u $^{40}$Ar +$^{197}$Au reactions. The results presented here demonstrate that hydrogen and helium were observed with high isotopic resolution, and the HBT correlation functions of light charged particles could be constructed from the obtained data.
\end{abstract}


\maketitle

\section{Introduction}\label{sec.I}

The two-particle small-angle correlation function at small relative momenta, originally developed by Hanbury Brown and Twiss (HBT)~\cite{hbt1956}, is a useful method for studying the space-time characteristics of the nuclear reaction zone and the emission order of particles in heavy ion reactions at intermediate beam energies~\cite{kim1991,verde2006,fdq2016,fdq2020}. 
In recent decades, the HBT correlations of intermediate mass fragments and light particles have been widely measured~\cite{lzy1996,hzy1997,xzg2006}, revealing many interesting space-time features of heavy ion reactions. As demonstrated in many experimental studies in nuclear physics over a wide energy range, the application of silicon strip detectors (SSDs) greatly improves the detector granularity and momentum resolution and has enabled the output of numerous novel and intriguing physical results ~\cite{hira2007,lassa2001,lq2018,dff2018,zgl2017,xxx2018,ws2020,wdx2020}. Similarly, more precise measurement of the HBT correlation function using this technique can be expected.

In addition, the isospin dynamics has attracted attention because it carries considerable information about the nuclear symmetry energy~\cite{lba2008,lba2019,zy2017,suj2020,yuh2020,wgf2020}, which is important not only in heavy ion reactions but also in processes involving dense stellar objects, including neutron star merging~\cite{ligo2017,ligo2018}. 
To determine the isospin migration rate, which reflects the effect of the symmetry energy, it is necessary to measure the emission rate as a function of time of light charged particles with different $N$/$Z$ values from the reactions. This measurement reveals the isospin chronology and requires high-resolution determination of both the position and energy to reconstruct the HBT correlation function, which must be done before the isospin-dependent emission time constant can be derived. For this purpose, position-sensitive detectors with high granularity and particle identification, for example, SSDs, are desired.

To experimentally measure the isospin chronology, we built the Compact Spectrometer for Heavy Ion Experiments (CSHINE). This paper discusses mainly the construction and performance of the SSD telescopes in CSHINE. After a brief introduction to the phase-I setup of CSHINE in Sect.~\ref{sec.II}, the performance of the SSD telescopes in the source test and beam experiment are presented in Sect.~\ref{sec.III}. Section~\ref{sec.IV} presents the $\alpha-\alpha$ correlation function obtained in the beam experiment and possible future applications. Section~\ref{sec.V} presents a summary.

\section{CSHINE detection system and beam experiment setup}\label{sec.II}

In phase I, two main types of detectors are installed on CSHINE: SSD telescopes for charged particle measurements and parallel plate avalanche counters (PPACs) for fission fragment measurements, which deliver the reaction geometry by folding angle reconstruction. In addition, three Au(Si) surface barrier telescopes are installed at large angles to measure the evaporated particles. Each SSD telescope is a three-layer detector with a single-sided SSD (SSSD) to measure $\Delta$E1 (layer 1), a double-sided SSD (DSSD) to measure $\Delta$E2 (layer 2), and a  $3\times3$ CsI(Tl) array for residual energy measurements (layer 3). Both the SSSD and DSSD are the BB7 type (2 mm strip width, 32 strips per side) from MICRON Company. The nominal thicknesses of these two layers of the SSD are 65 and 1500 $\mu$m, respectively. Owing to the 2 mm strip width, the angular resolution in the beam experiment, which is a key parameter for the measurement of the small-angle correlation function, is better than $1^\circ$. 

\begin{figure}[!htb]
\includegraphics[width=1.3\hsize]{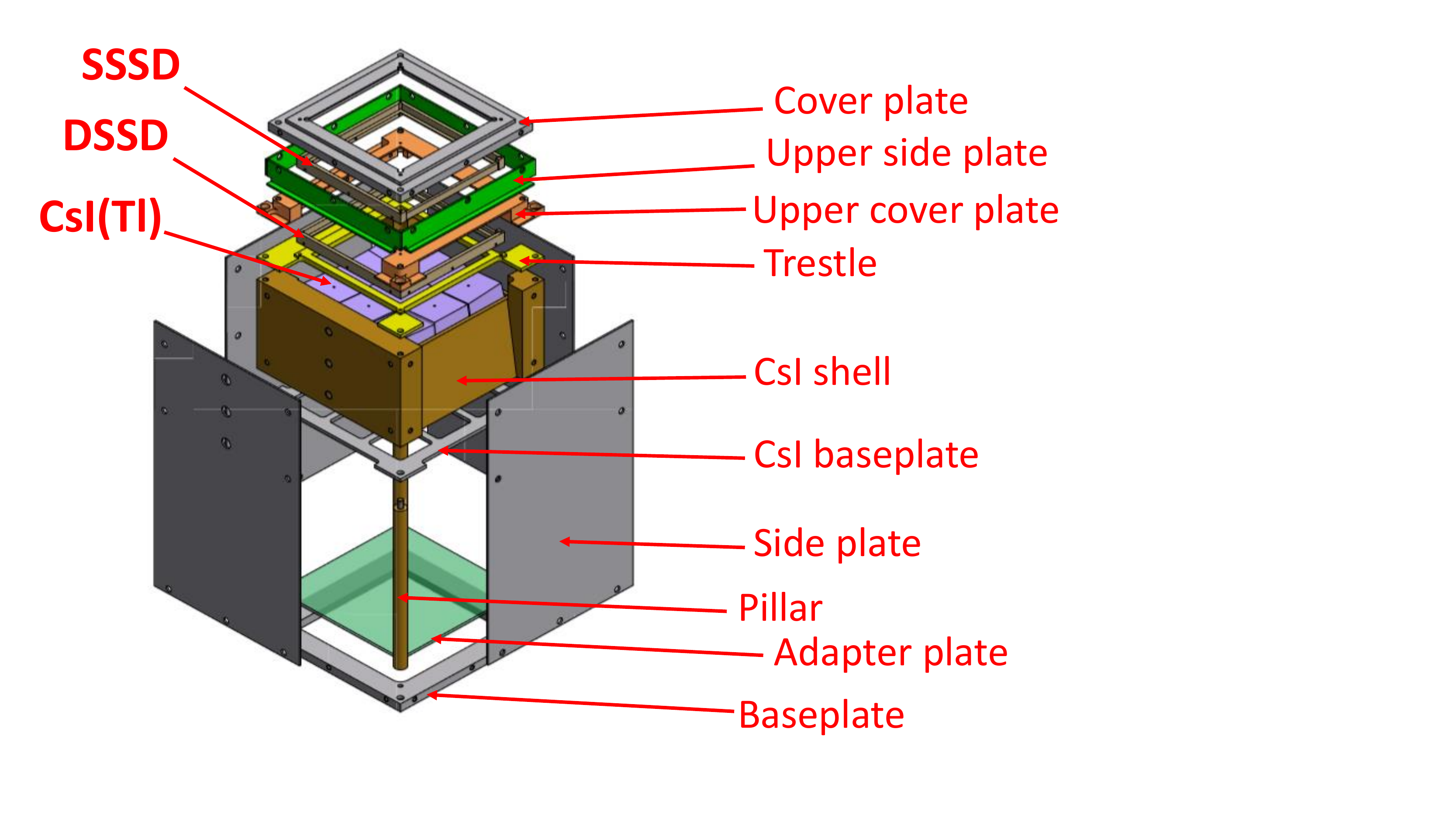}
\caption{(Color online) Structural framework of SSD telescope.}
\label{fig1}
\end{figure}

Each CsI(Tl) crystal is a square pyramid with dimensions of $23\times23$ mm$^2$ on the front side and $27\times27$ mm$^2$ on the rear side and a height of 50 mm. The well-polished CsI(Tl) crystals are wrapped with Teflon for good light reflection. The entrance face is covered by 2-$\mu$m-thick single coat of aluminized Mylar foil, and the other sides are packed with black tape for tight light shielding. At the rear end of each CsI(Tl) crystal, there is no wrapping material, and the crystal is directly coupled to a S3204 photodiode (Hamamatsu) by BC-630 optical grease (Saint-Gobain). Each SSD telescope is mounted in an aluminum frame. The structural framework of each SSD telescope is shown in Fig. ~\ref{fig1}.

Each SSD has an energy resolution of approximately 1\% (full width at half-maximum, FWHM) for 5.15 MeV $\alpha$ particles,
and the CsI(Tl) detector has an energy resolution of approximately 20\% (FWHM) for this $\alpha$ source. The other type of detector, the PPAC, is a gas detector with a large sensitive area ($240\times280$ mm$^2$), a two-dimensional position resolution of approximately 1.35 mm, and a time resolution of less than 300 ps. These PPAC detectors were described in detail in Ref.~\cite{wxl2019}.
\begin{figure}[!htb]
\includegraphics[width=1.1\hsize]{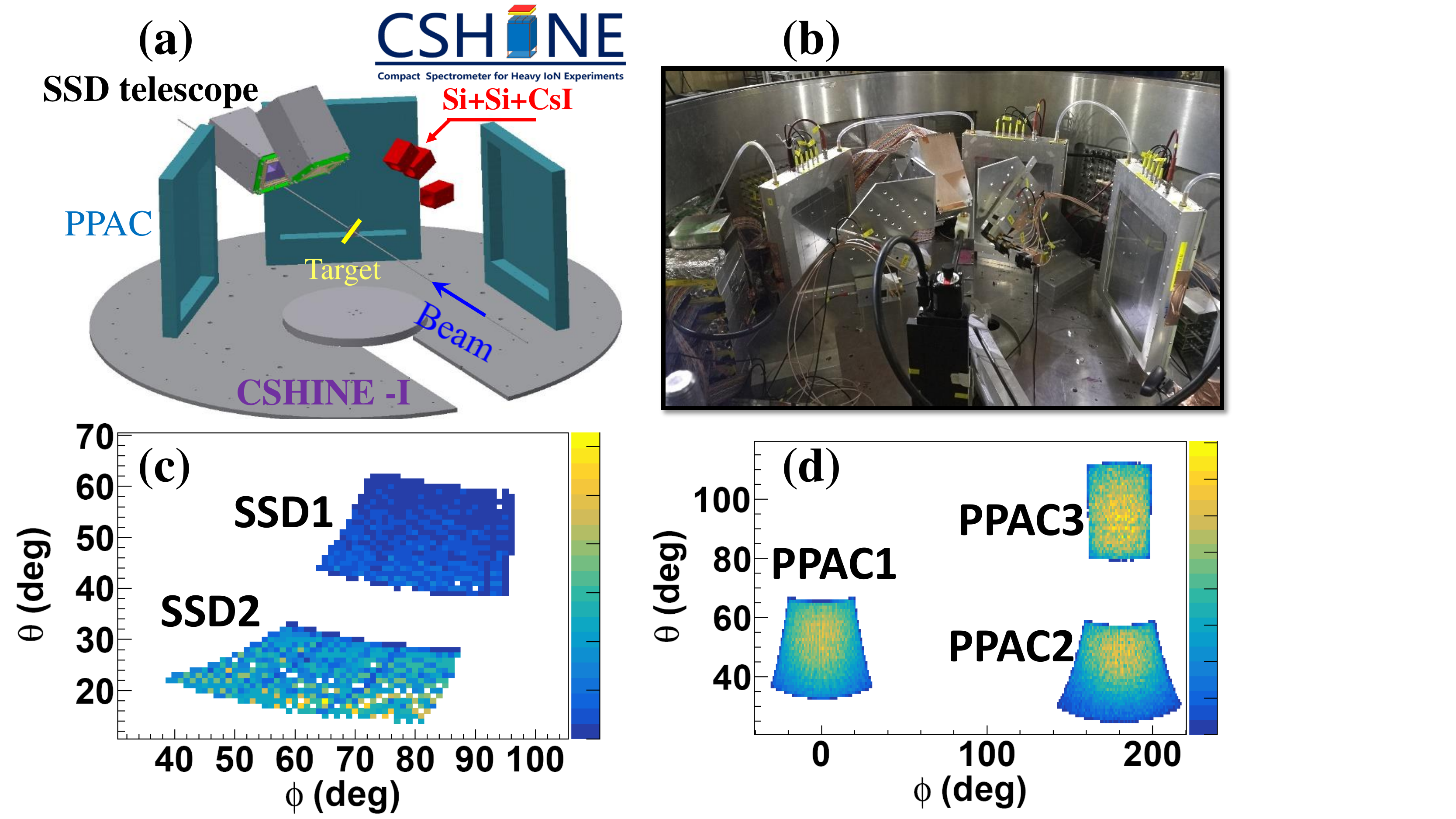}
\caption{(Color online) CSHINE detection system in phase I. (a) Sketch of CSHINE detection system. (b) Photograph of CSHINE detection setup in actual experiment. (c) Angular coverage of SSD telescopes. (d) Angular coverage of PPACs.}
\label{fig2}
\end{figure}

\begin{table}[!htb]
\caption{Experimental geometry parameters}
\label{tab:Experiment-geometry-parameters}
\begin{tabular*}{8cm} {@{\extracolsep{\fill} } llllr}
\toprule
Detector & $d$ (mm) & $\theta$ ($^\circ$) & $\phi$ ($^\circ$) & $S$ (mm$^2$) \\
\midrule
SSD-Tele1  & 161.9  & 50.7 & 81.7  & $64\times64$  \\
SSD-Tele2  & 221.9  & 22.3 & 64.5  & $64\times64$   \\
PPAC1      & 427.5  & 50   &  0    & $240\times280$   \\
PPAC2      & 427.5  & 40   &  180  & $240\times280$    \\
PPAC3      & 427.5  & 95   &  180  &$240\times280$    \\
\bottomrule
\end{tabular*}
\end{table}

\begin{figure*}[!htb]
\includegraphics[width=1.0\hsize]{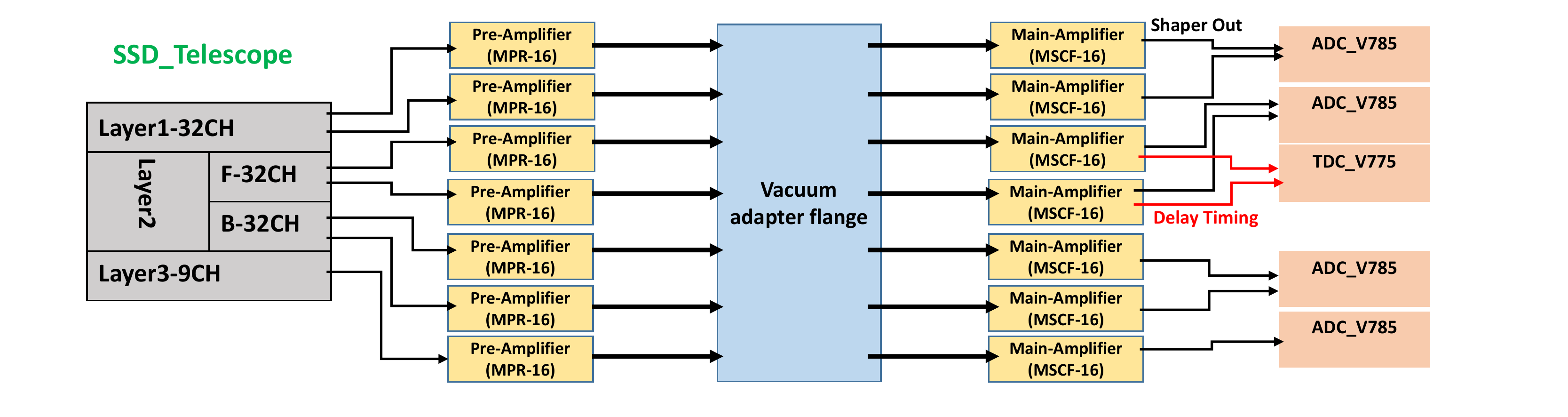}
\caption{Schematic of SSD telescope signal data flow.}
\label{fig3}
\end{figure*}

\begin{figure*}[!htb]
    \includegraphics[width=1.0\hsize]{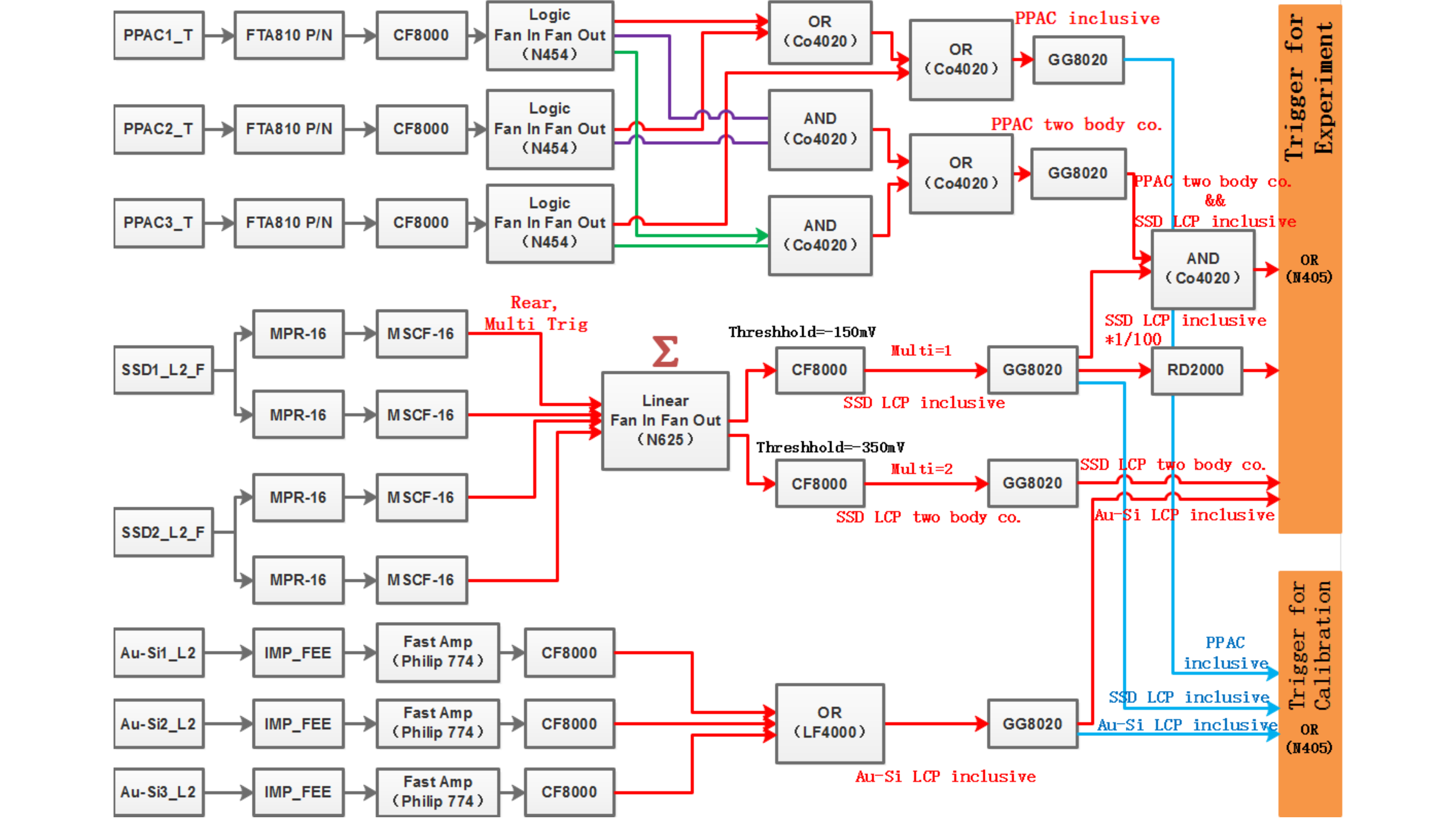}
    \caption{Diagram of CSHINE trigger system.}
    \label{fig4}
\end{figure*}

The beam experiment was performed at Radioactive Ion Beam Line I in Lanzhou (RIBLL1) at the Heavy Ion Research Facility in Lanzhou, China. A $^{197}$Au target with an areal density of 1 mg/cm$^2$ was bombarded with a 30 MeV/u $^{40}$Ar beam. 
Although the entire CSHINE detection system contains six SSD telescopes and three PPACs, in phase I, two SSD telescopes and three PPACs were mounted for the beam experiment. 
The SSD telescopes were installed at forward angles to detect light particles at mid-rapidity in coincidence with the fission fragments measured by the three PPACs. 
A graph illustrating the CSHINE detection system geometry in the first experiment is shown in Fig.~\ref{fig2}. 
Table 1 lists the distance $d$ from the center of each detector to the target, the polar angle $\theta$, the azimuthal angle $\phi$, and the sensitive area $S$ of the two SSD telescopes and three PPACs in the experiment.

Each SSD telescope consists of one SSSD, one DSSD, and one CsI(Tl) hodoscope containing $3\times3$ CsI(Tl) crystal units.  
The SSSD has 32 readout strips on the front side, whereas the DSSD has 32 readout strips on two sides perpendicular to each other. A total of 96 individual silicon strips and 9 CsI(Tl) crystals require readout electronics. In a simple readout scheme, a compact pre-amplifier (MPR-16, Mesytec) and main amplifier (MSCF-16, Mesytec) were used. The pulse height of the signals carrying the energy information was digitized by a 12-bit analog-to-digital converter (ADC) (V785, CAEN), and the timing information was digitized by a 12-bit time-to-digital converter (V775, CAEN). Finally, all the digital signals were recorded on a hard disk event-by-event using a data acquisition system based on the VME standard. Fig.~\ref{fig3} presents the data flow of the SSD telescopes.

The CSHINE trigger system was designed for both the beam experiment and calibration. Fig.~\ref{fig4} presents a diagram of the trigger circuit. The timing signals of the PPACs were discriminated by a CF8000 module and logically calculated by a CO4020 module to generate the PPAC inclusive signals and PPAC two-body coincidence signals. 
The logic hit signals of the SSD telescopes were obtained by the front side of the DSSD ($\Delta E2$) and discriminated by the MSCF-16 module, which generated an analog Multi-Trig signal proportional to the number of fired strips in this module (16 channels) and sent it to the output on the rear panel. The Multi-Trig signals from all the MSCF-16 modules then proceeded to an analog fan-in module (N625, CAEN) to form a summation signal, the height of which represents the total multiplicity of the fired strips in the SSD telescopes. Then, both inclusive and exclusive logic signals were generated by the discriminator (CF8000 module) at different threshold settings. The logic hit signals of the Au(Si) telescopes were generated from the Au(Si) surface barrier detector in layer 2 and logically added by an LF4000 module to create the inclusive signal of the Au(Si) telescopes. After the width and time delay of every logic signal were adjusted, the trigger signal for the experiment was constructed. The trigger signal contains the coincidence of the PPAC two-body and SSD light charged particle inclusive events and the SSD two-body events for data acquisition in the beam experiment. In addition, the inclusive trigger for each detector was also constructed and was optionally turned on for detector calibration before or after beam data acquisition.

\section{Performance of SSD telescopes and particle identification}\label{sec.III}
\subsection{Energy resolution of SSDs and CsI(Tl) in $\alpha$ source test}\label{sec.A}

\begin{figure}[!htb]
\includegraphics[width=1.2\hsize]{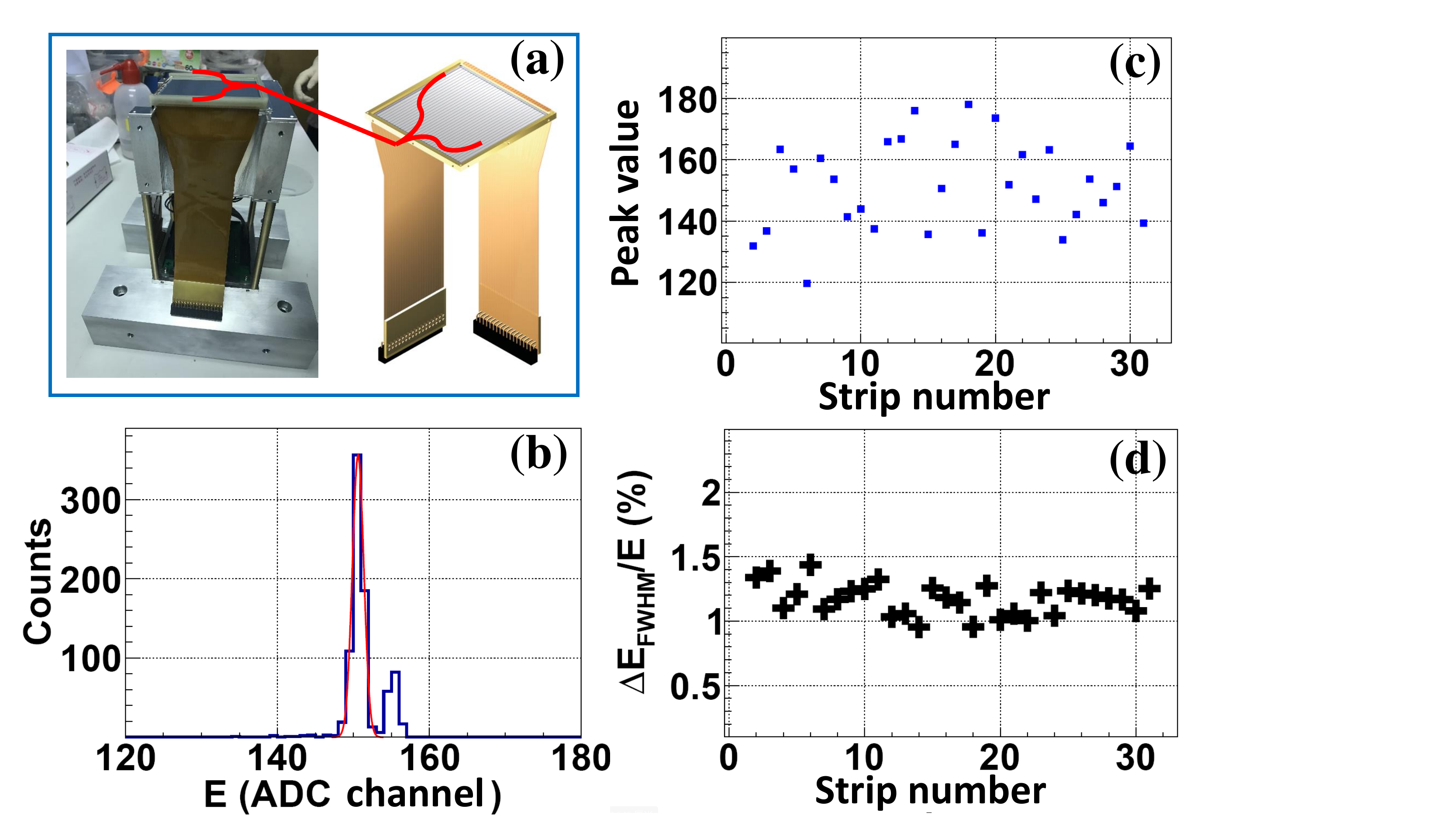}
\caption{(Color online) Energy resolution of DSSD with $\alpha$ source. (a) Photograph and sketch of DSSD. (b) Energy spectrum of $\alpha$ source in a single strip. (c) Position of the main peak of the $\alpha$ source for all strips in this telescope. (d) Energy resolution of all strips in this telescope.}
\label{fig5}
\end{figure}

Figure~\ref{fig5} presents the structure of the DSSD and its performance during a test using a 5.15 MeV $\alpha$ source for SSD telescope 1. Figure~\ref{fig5} (a) presents a photograph of the telescope during assembly. Fig.~\ref{fig5} (b) shows the energy spectrum of the $\alpha$ source for a single strip. Because the energy loss of light charged particles in the DSSD will be much larger in the beam experiment than that of 5.15 MeV $\alpha$ particles, low amplification was used so that the $\alpha$ peak appeared in low channels. The two peaks at 5.15 and 5.49 MeV ($^{239}$Pu and $^{241}$Am, respectively) can be separated clearly. Figure~\ref{fig5} (c) shows the position of the main peak of the $\alpha$ source for all the strips in this telescope. The energy responses are distributed in a range corresponding to 120--180 ADC channels and show significant variation. Although the variation in the peak position is quite large, the energy resolution of all the strips remains uniform at approximately 1\% (FWHM), as shown in  Fig.~\ref{fig5} (d). The performance of this DSSD, which is installed at a large angle, remains stable before and after the beam experiment. Note, however, that the performance of the other DSSD, which is located near the beam axis and receives more irradiation, gradually deteriorated, as reflected by the increasing leakage current. To solve this problem, our experience with a recent experiment indicates that it is necessary to cool the detector using an alcohol refrigerator so that both the leakage current and the resolution remain unchanged during the beam experiment. 

\begin{figure}[!htb]
\includegraphics[width=1.2\hsize]{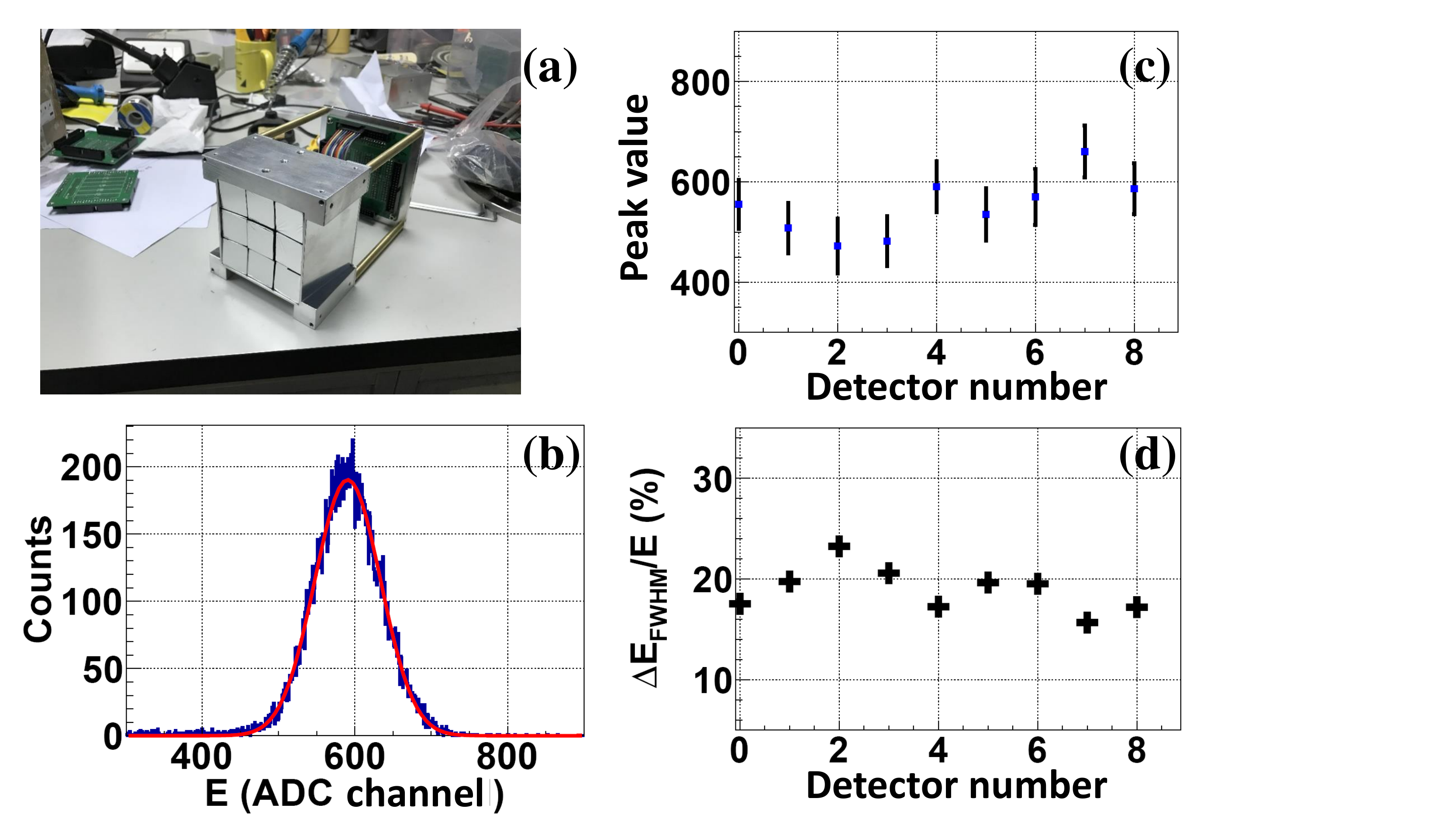}
\caption{(Color online) Energy resolution of CsI(Tl) array with $\alpha$ source. (a) Photograph of CsI(Tl) array. (b) Energy spectrum of $\alpha$ source in a single CsI(Tl) crystal. (c) Position of main peak of $\alpha$ source for all CsI(Tl) crystals in this telescope. (d) Energy resolution of all CsI(Tl) crystals in this telescope.}
\label{fig6}
\end{figure}

Figure~\ref{fig6} presents the same results, but for the $3\times3$ CsI(Tl) array and the 5.15 MeV $\alpha$ source during offline calibration. The energy resolution is only approximately 20\% (FWHM) [Fig.~\ref{fig6} (d)]. One possible reason for the low energy resolution of the CsI(Tl) array is that the $\alpha$ particle is stopped at a depth of 27 $\mu$m in the CsI(Tl) material (as calculated by LISE++~\cite{lise2016}), which is much less than the total length of the CsI(Tl) crystal (50 mm). The light response  is significantly suppressed, and the light transport efficiency is low. During the beam experiment, the energy resolution for particles with a larger stopping depth will be much higher (see text below).

\subsection{Signal sharing in SSD}\label{sec.B}

\begin{figure}[!htb]
    \includegraphics[width=1.0\hsize]{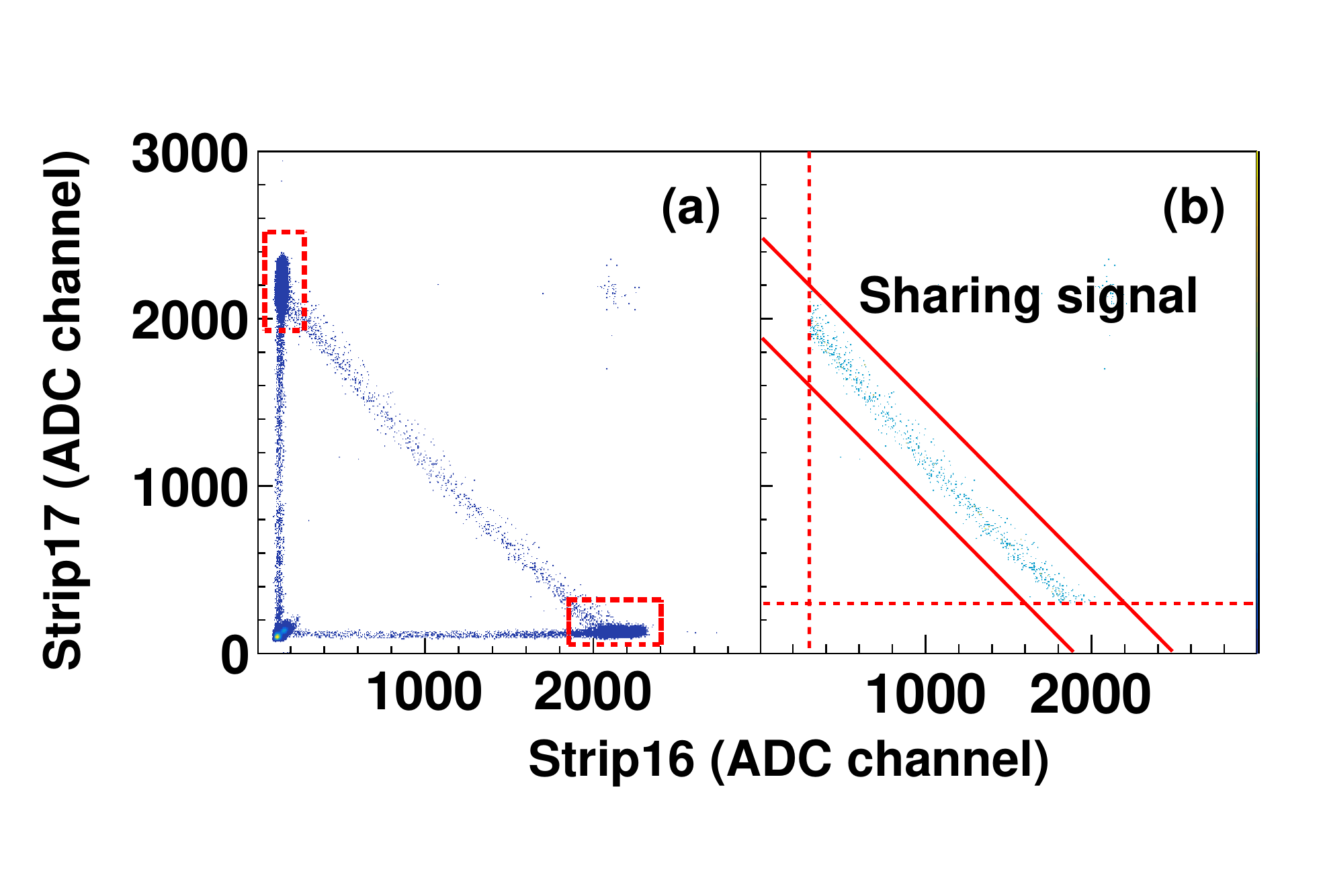}
    \caption{(Color online) Energy correlation spectrum between two neighboring strips of the 65 $\mu$m SSSD for all (a) and only signal-sharing (b) events.}
    \label{fig7}
\end{figure}

Signal sharing is not negligible when the SSDs are used because the interstrip distance is only 0.1 mm. 
When an incident particle strikes one strip in the detector, the neighboring strips are likely to deliver a signal including a certain portion of the total charge produced by ionization. The proportion of signal sharing events affects the efficiency of particle identification. Taking the 65 $\mu$m SSSD as an example, Figure~\ref{fig7} (a) shows the energy correlation spectrum in two neighboring strips in the  $\alpha$ source test. Most events are recorded by one strip; the amplitude of the signal is located at the full energy position, and the signal of the neighboring strip is located in the pedestal, as indicated by the two rectangles in the plot.  In addition, some events are located in the center of the coincident plot connecting the two main peaks, where the sum of the energy of the two strips has a constant value corresponding to the full energy of the incident $\alpha$ particle. Figure~\ref{fig7} (b) presents the distribution of only signal-sharing events. 

\begin{figure}[!htb]
\includegraphics[width=1.0\hsize]{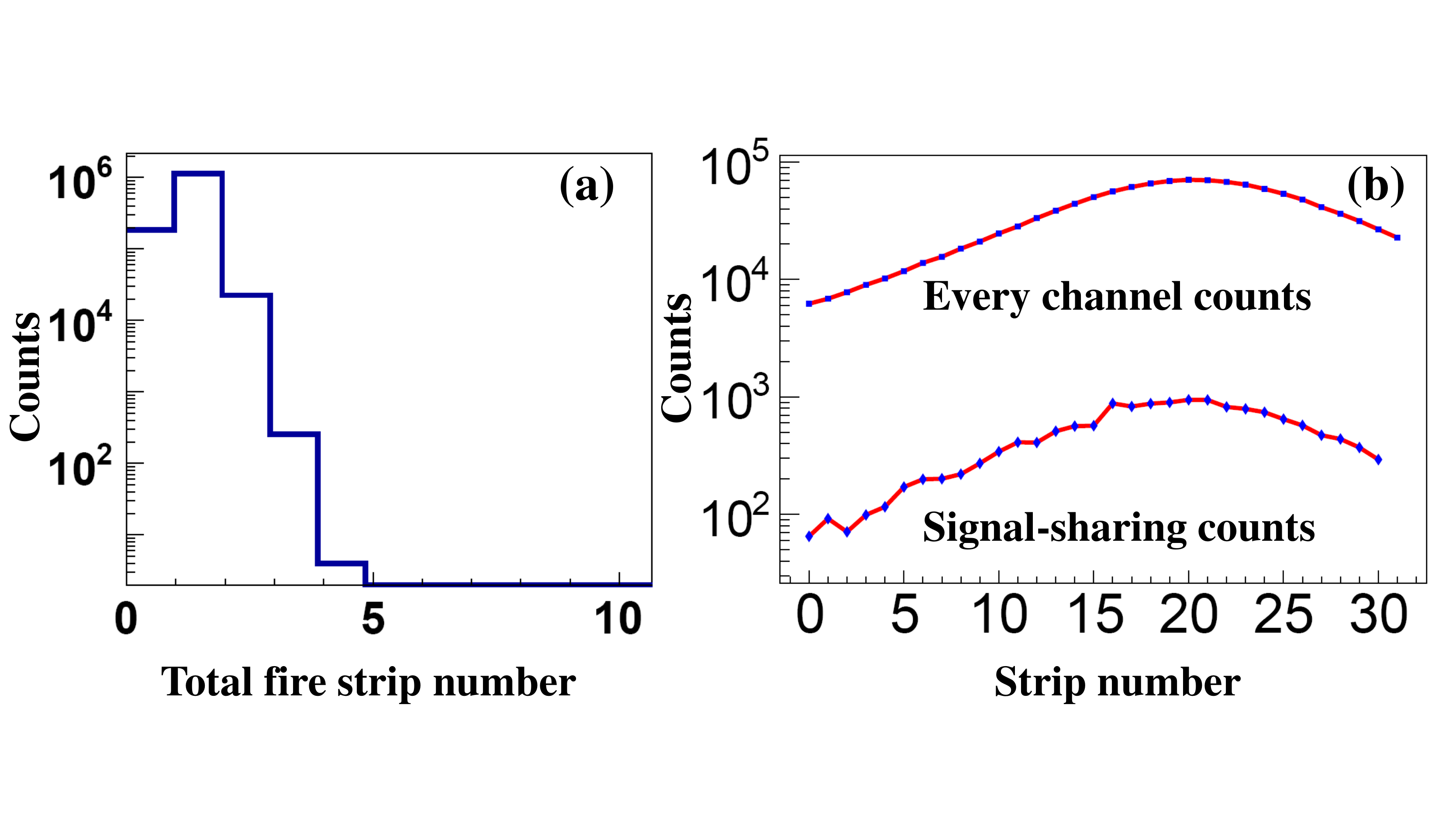}
\caption{Multiplicity $M_\text{str}$ of the firing strips after pedestal cut (a) in $\alpha$ test. Full energy counts and signal-sharing counts in each strip (b).}
\label{fig8}
\end{figure}

Figure~\ref{fig8} (a) presents the multiplicity of the firing strips in the $\alpha$ test after a pedestal cut, $M_\text{str}$, is applied. Most of the events satisfy $M_\text{str}=1$ or 2, and the latter are contributed mainly by signal sharing in neighboring strips. By counting the events above the pedestal cut in the individual strips and the signal-sharing events of neighboring strips, the ratio of signal-sharing events on each strip is found to be less than 1\%, as shown in Fig.~\ref{fig8} (b). This ratio is used as a reference value in the data analyses of the particle identification efficiency. 

\subsection{Particle identification}\label{sec.C}

\begin{figure}[!htb]
\includegraphics[width=1.1\hsize]{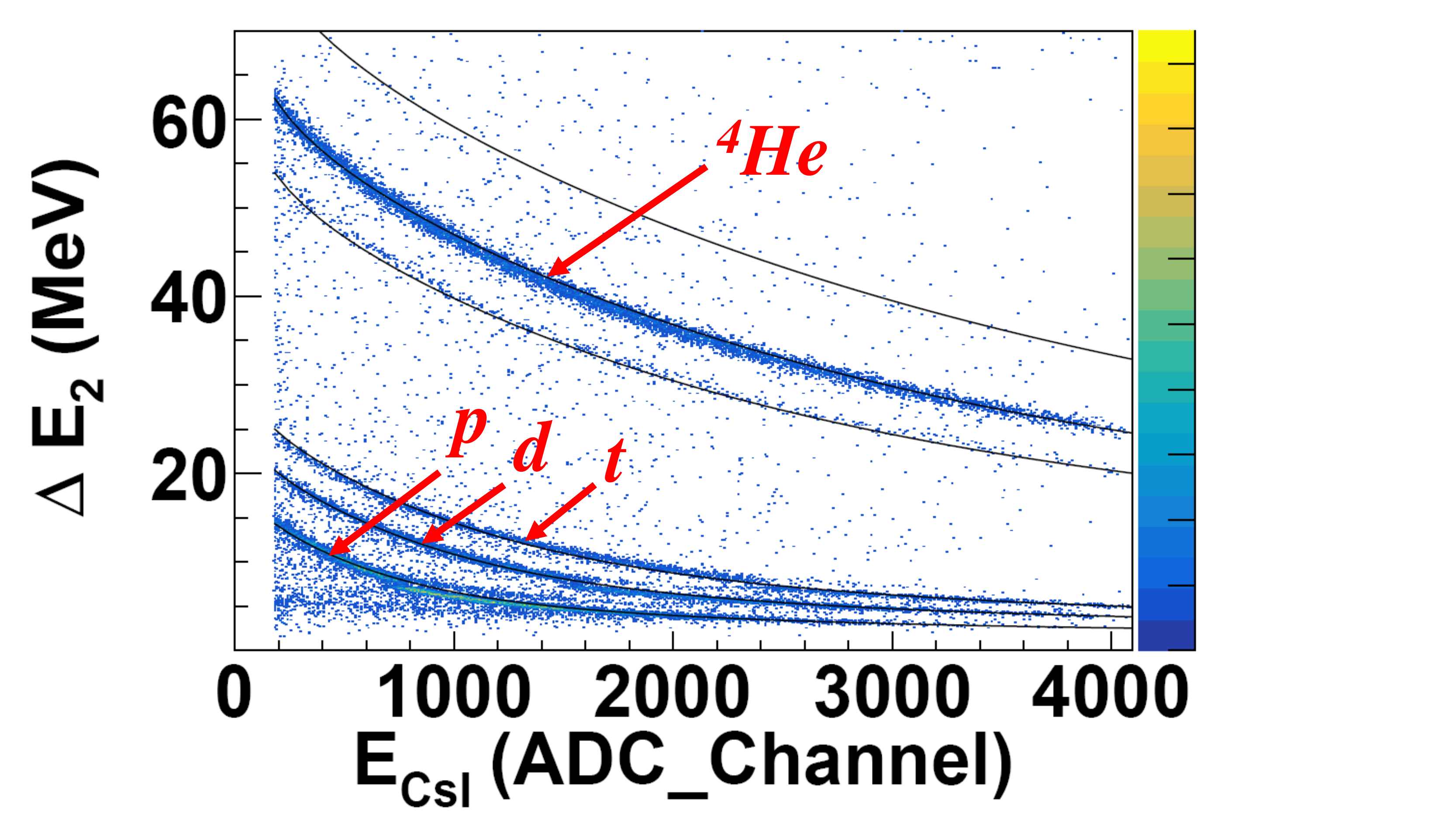}
\caption{(Color online) Two-dimensional scattering plot of energy loss $\Delta E$ in SSD layer 2 versus the deposited energy $E$ in one CsI(Tl) unit.}
\label{fig9}
\end{figure}

\begin{figure}[!htb]
\includegraphics[width=1.0\hsize]{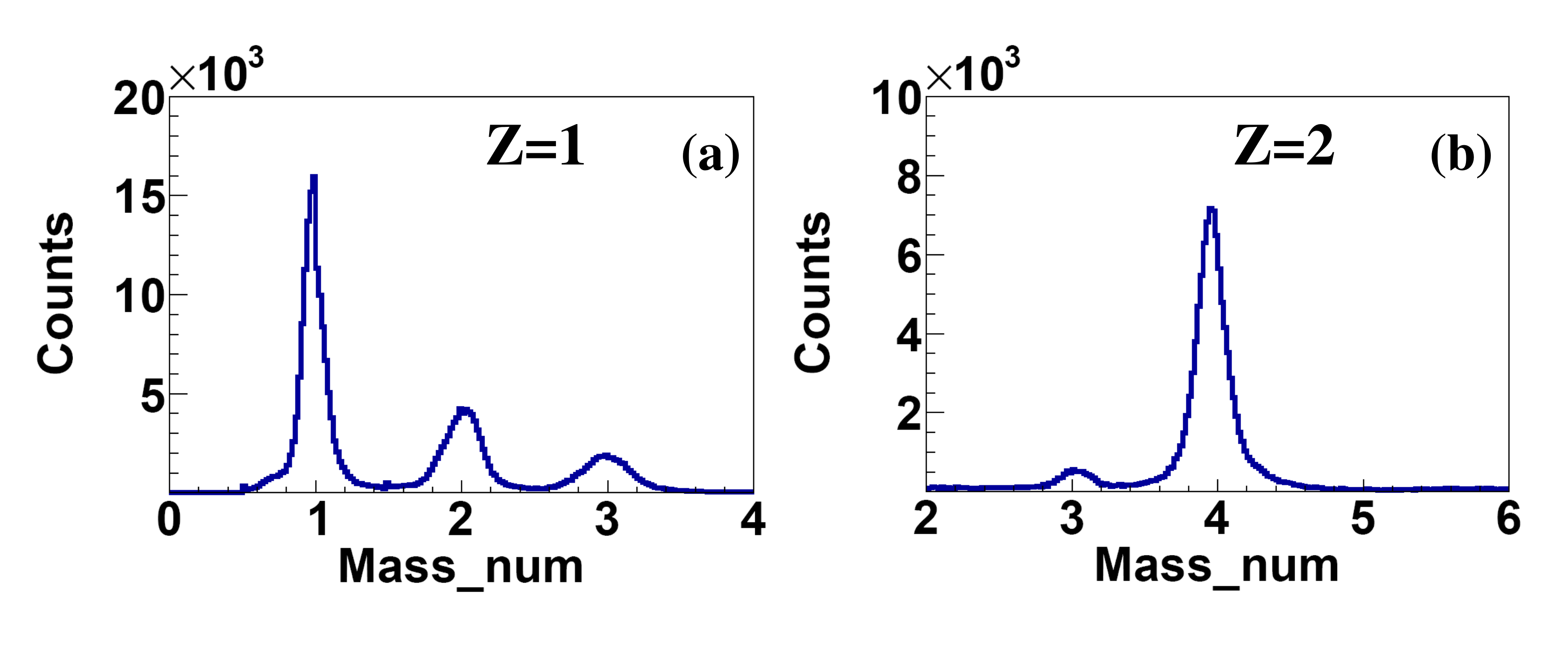}
\caption{Mass spectra of $Z=1$ (a) and $Z=2$ (b) isotopes.}
\label{fig10}
\end{figure}

Figure~\ref{fig9} shows the scattering plot of the energy loss $\Delta E$ in the DSSD transmission detector in layer 2 versus the energy $E$ deposited in the CsI(Tl) crystal in the 30 MeV/u $^{40}$Ar +$^{197}$Au reaction. In the beam experiment, we focused on the measurement of light charged particles; therefore, the energy in layer 1 is very small and is not presented here. The spectrum shows that the isotopes of $Z=1$ and 2 elements are clearly separated in the entire measuring energy range. The mass resolution was obtained using the linearization method. Briefly, the band of each isotope can be described by a trend curve obtained by fitting manually chosen marker points with a 14-parameter function. Each curve represents the mass of the corresponding isotope. Then, during data sorting, for each particle with a given ($\Delta E$ - E) value, the vertical distances between the point ($\Delta E$ - E) and the two neighboring curves define the experimental mass number. Figure~\ref{fig10} presents the mass number spectra for $Z=1$ (a) and $Z=2$ (b) isotopes. By fitting each peak with a Gaussian function, the mass resolution can be derived. As shown in Fig.~\ref{fig10} (b), a mass resolution of $\Delta M=0.1$ can be obtained for $\alpha$ particles. 
The acquisition of a full scattering plot with high efficiency requires a sophisticated pattern recognition algorithm. The entire procedure for determining the tracks using the event-by-event logic for events with a total multiplicity of $M=1$ and 2 will be reported elsewhere.

\begin{figure}[!htb]
\includegraphics[width=1.1\hsize]{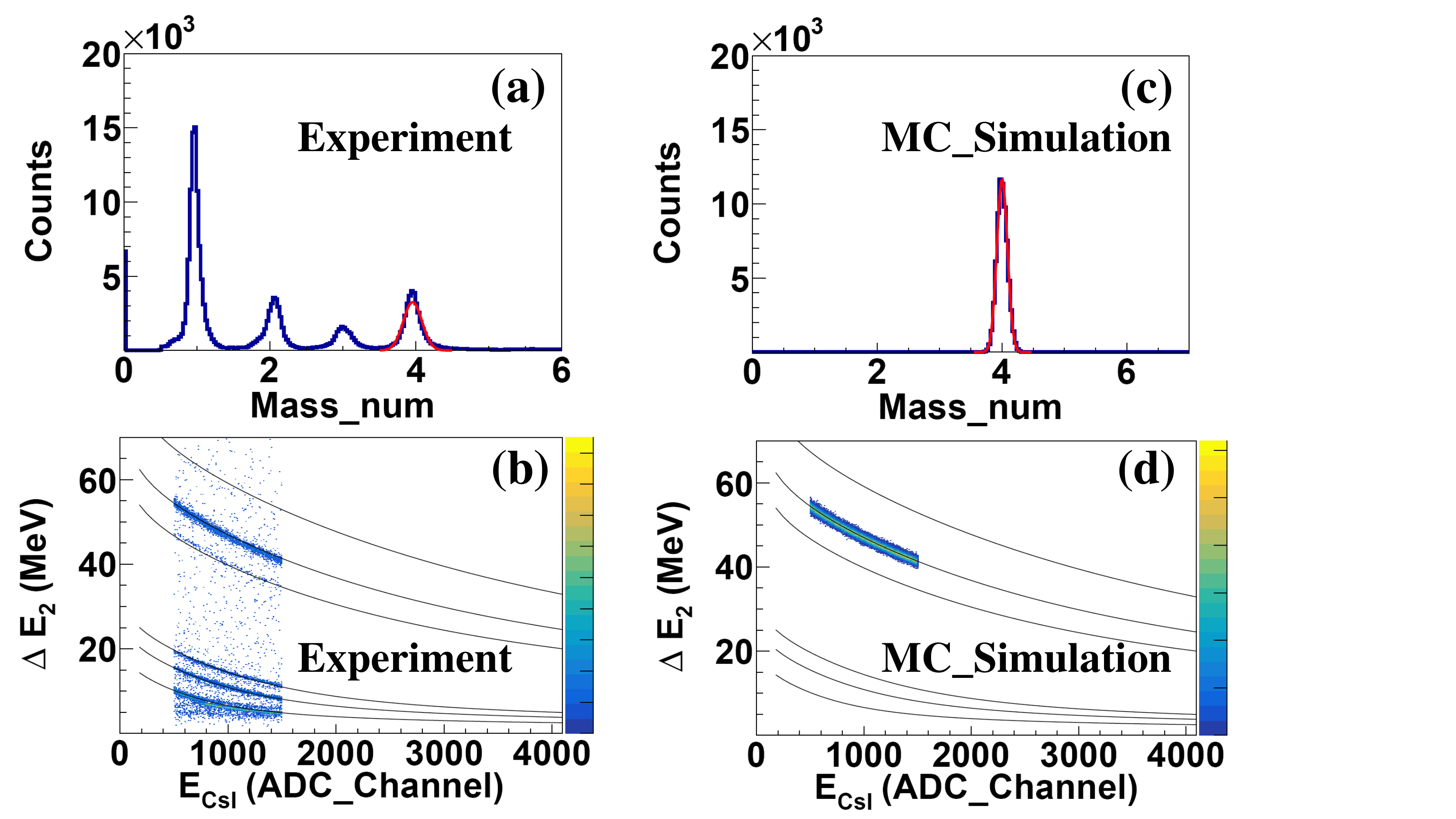}
\caption{(Color online) Comparison of experimental and MC simulation results in an $E$ range from 500 to 1500 ADC channels. (a) Mass distribution from experimental data and Gaussian fit of peak for $A=4$, (b) experimental $\Delta E-E$ scattering plot, (c) mass distribution generated by MC simulation with energy resolution of 1.1\% (1$\sigma$) for $\Delta E$ and 2\% for $E$ (1$\sigma$) and Gaussian fit of peak for $A=4$, (d) simulated $\Delta E-E$ scattering plot.}
\label{fig11}
\end{figure}

From the mass resolution, one can deduce the energy resolution of the CsI(Tl) detector, which is expected to be better in the beam experiment than in the $\alpha$ source test because the stopping depth of the 5.15 MeV $\alpha$ particles is very small at the surface.  
Because the data points are located around the trend curve, as shown in Fig.~\ref{fig9}, the bandwidth is attributed to the finite energy resolution of the SSDs and CsI(Tl) array. To extract the energy resolution of the CsI(Tl) array and the DSSD, a Monte Carlo simulation was conducted.  By varying the energy resolution of both units [$\Delta E$ for the DSSD and $E$ for CsI(Tl)], different broadening of the mass spectrum can be simulated, and the mass resolution can be derived. This procedure can be performed in a divided range on the $\Delta E-E$ plot. Fig.~\ref{fig11} presents the experimental mass spectrum (a) and $\Delta E-E$ plot (b), which are compared to the MC simulations in (c) and (d), respectively. 
The $\sigma$ value from the Gaussian fit of the $\alpha$ mass distribution is used to estimate the energy resolution.
By comparing the width of the simulated mass peak with the experimental results, one can estimate the energy resolutions of both units, which are found to be correlated. 

\begin{figure}[!htb]
\includegraphics[width=0.9\hsize]{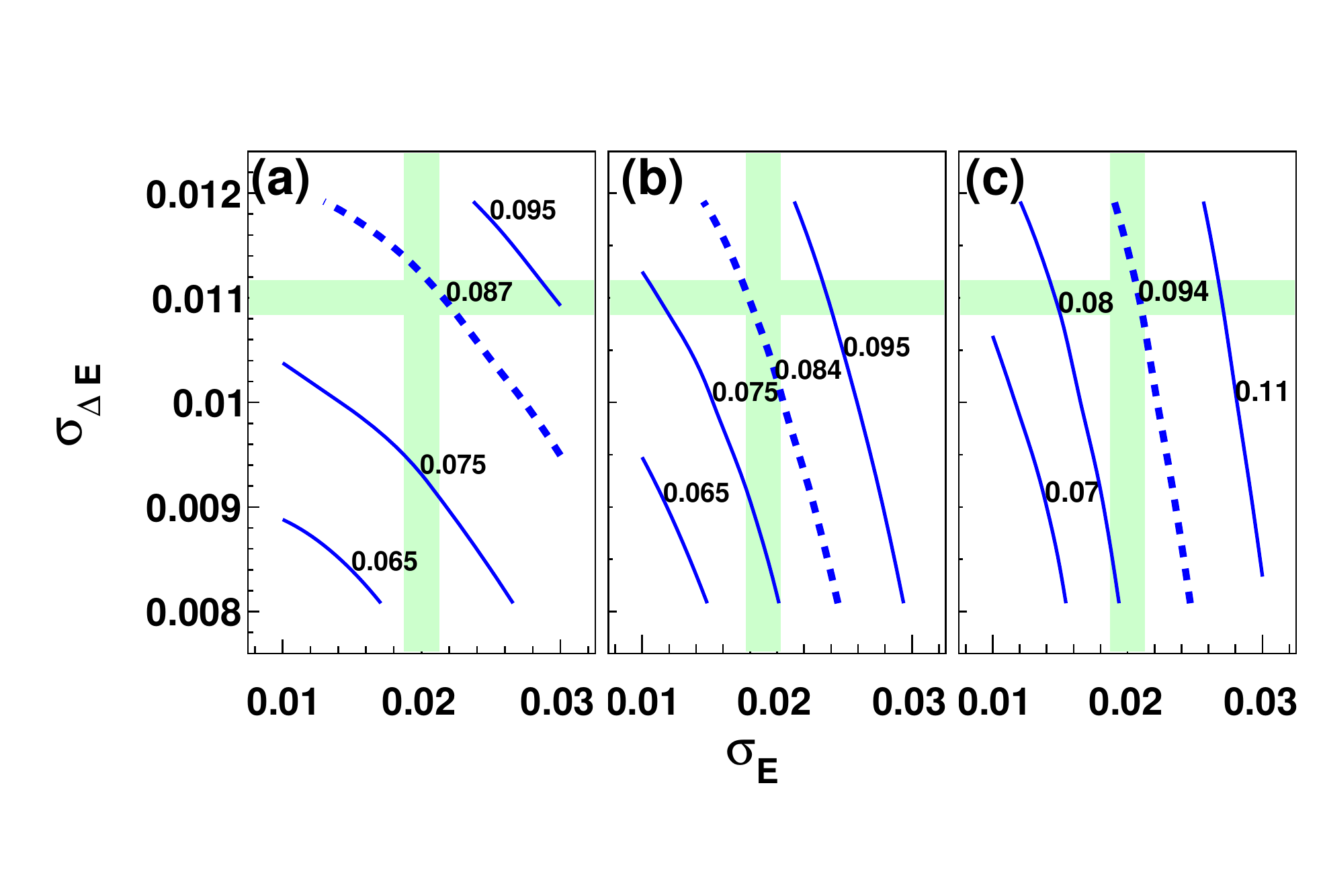}
\caption{(Color online) Monte Carlo simulation results of the total mass resolution as a function of energy resolution $\Delta E$ for DSSD and $E$ for CsI(Tl) in three $E$ ranges (in units of ADC channels) of $500-1500$ (a), $1500-2500$ (b), and $2500-3500$ (c). The number near the curve is the total mass resolution obtained in the simulation. The dashed curves correspond to the mass resolution obtained from the experimental data in each interval.}
\label{fig12}
\end{figure}

 Figure~\ref{fig12} presents the contour of the mass resolution as a function of the resolution $\Delta E$ of the DSSD and $E$ of the CsI(Tl) array in three $E$ ranges. As both parameters increase, the variation in the mass distribution increases. The slopes of the contour curves are different because of the bending of the $\Delta E-E$ band, as shown in Fig.~\ref{fig9}. Thus, the energy resolutions of the DSSD and CsI(Tl) contribute different weights to the total mass resolution in each range. The dashed curves in the plot represent the approximate total experimental mass resolution in the selected $E$ range. A comparison of Figs.~\ref{fig12} (a), (b), and (c) reveals that the experimental energy resolution of the DSSD, including the intrinsic energy resolution and hit pattern recognition error, is approximately $\sigma_{\Delta E}=1.1\%$ (1 $\sigma$), which is consistent with the result of the $\alpha$ source test, whereas the energy resolution of CsI(Tl) in the beam experiment is approximately $\sigma_{E}=2\%$ (1$\sigma$), as indicated by the green bands. These results for the energy resolution of the SSD and CsI(Tl) are essential for physical analyses of the HBT correlation function.

\subsection{CsI(Tl) energy calibration}\label{sec.D}

The calibration of the CsI(Tl) unit is an important issue in the application of SSD telescopes~\cite{dell2019}. The energy response of the SSD is linear and independent of the particle species, and it can be calibrated using an $\alpha$ source and a precision pulse generator. However, the signal response of the CsI(Tl) scintillators is nonlinear and depends on the particle charge and mass. Ideally, the absolute energy calibration of CsI(Tl) detectors requires various particle species with well-defined charges, masses, and energies, which can be obtained using the primary beam delivered by an accelerator.

\begin{figure}[!htb]
\includegraphics[width=1.0\hsize]{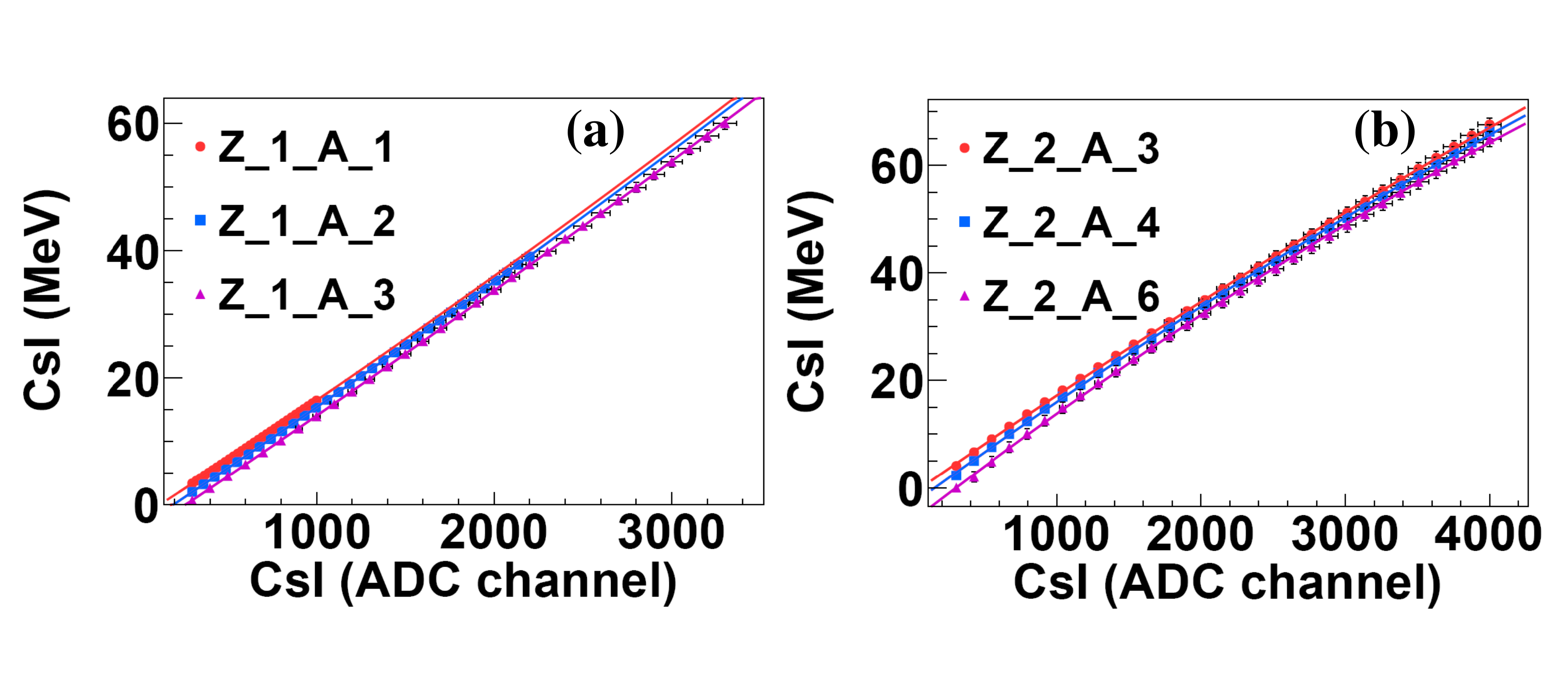}
\caption{(Color online) CsI(Tl) energy calibration for $Z=1$ (a) and $Z=2$ (b) isotopes.}
\label{fig13}
\end{figure}

However, because accelerator time is costly, we use a simpler method that is widely  employed. The CsI(Tl) signal responses are calibrated with the deposited particle energies using the $\Delta E-E$ plot. Because the SSD is well-calibrated, from the plot one can read multiple points on the curve following each isotope, where $\Delta E$ is calibrated in MeV and $E$ is calibrated using the raw ADC channel number. Given the nominal value of the thickness of the DSSD from the manufacturer and $\Delta E$ (MeV), the total energies and residual energies $E$ (MeV) in the CsI(Tl) can be computed for all points using the  LISE++ code \cite{lise2016}. The relationship between the ADC channel number and the energy deposited (MeV) in CsI(Tl) can be established. Figure~\ref{fig13} presents the energy calibration of $Z=1$ isotopes (a) and $Z=2$ isotopes (b) for CsI(Tl). The response of CsI(Tl) is nonlinear and moderately dependent on the charge and mass of the light charged particles. Moreover, a change in the nominal thickness of the DSSD by a few percent has a very insignificant effect on the CsI(Tl) calibration.

\section{Application of SSD telescopes and future uses of CSHINE}\label{sec.IV}

SSD telescopes have been used mainly to measure the HBT correlation functions of light charged particles in the reaction 30 MeV/u $^{40}$Ar + $^{197}$Au. The $\alpha-\alpha$ correlation function can be used to check the performance and calibration of the detector. There are three peaks in the relative momentum spectrum of $\alpha$ pairs, which originate mainly from the decay of the resonant states of $^8$Be and $^9$Be. The malfunctioning of the telescope would result in an incorrect $\alpha-\alpha$ correlation function.

\begin{figure}[!htb]
\includegraphics[width=1.0\hsize]{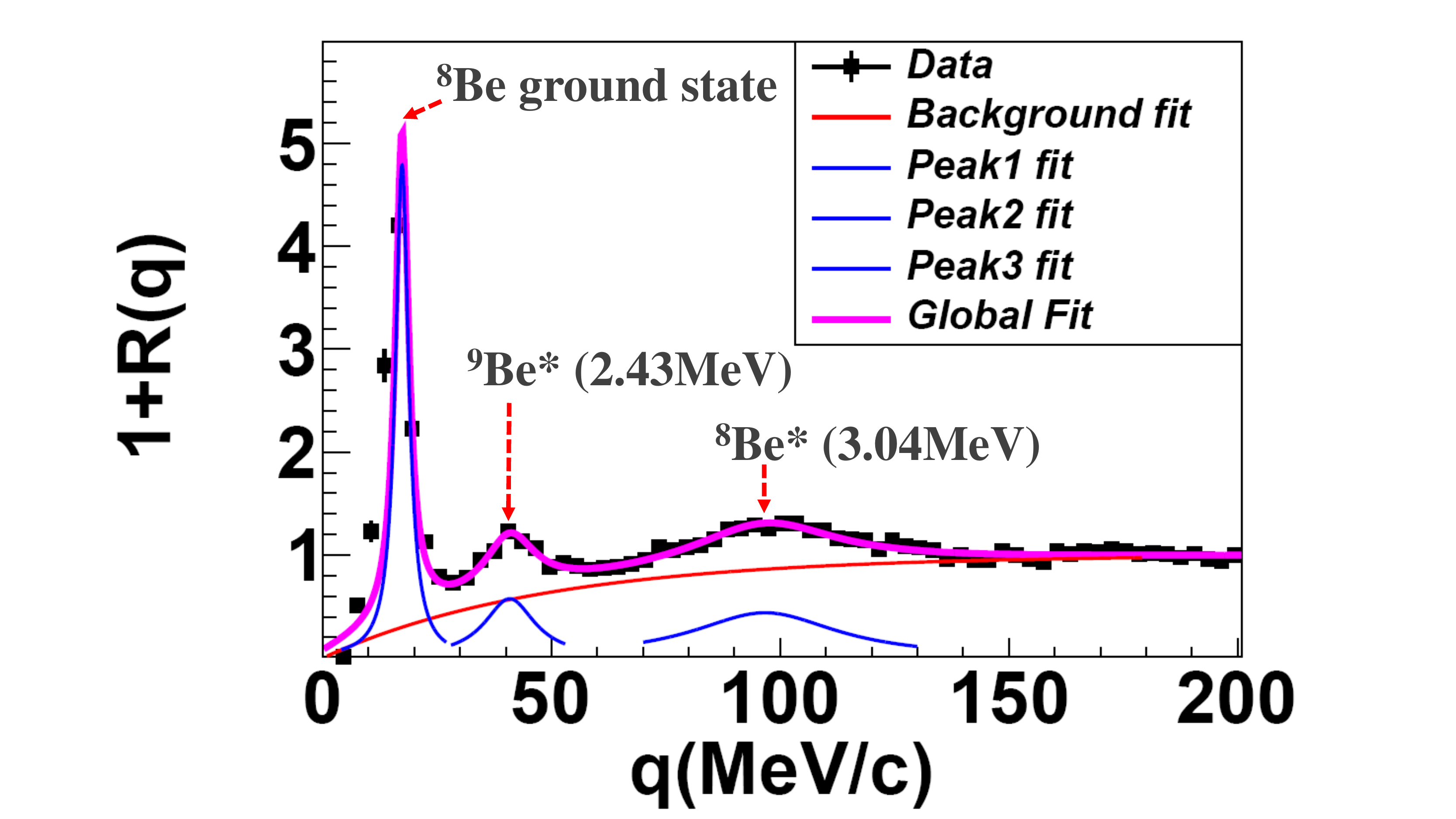}
\caption{(Color online) $\alpha$ – $\alpha$ correlation function for 30 MeV/u $^{40}$Ar + $^{197}$Au reaction.}
\label{fig14}
\end{figure}

Figure~\ref{fig14} shows the $\alpha-\alpha$ correlation function obtained by the two SSD telescopes. Various peaks are expected to appear in the correlation function. The peaks at 20 and 100 MeV/$c$ correspond to the decay of the unstable ground state and 3.04 MeV excited state of $^8$Be, respectively, whereas the peak at 50 MeV/$c$ corresponds to the decay of the 2.43 MeV excited state of $^9$Be~\cite{poch1987}. The positions of the three peaks are in agreement with the theoretical predictions, suggesting that the performance of the SSD telescopes is acceptable.

Isospin dynamics are very important for understanding the effect of nuclear symmetry energy in nuclear collisions. 
By using the SSD telescopes in CSHINE, the correlation functions of different particle pairs can be measured to extract the emission time constant and emission hierarchy of the species. 
In addition, the isotope-resolved particles in coincidence with the fission fragments measured using the PPACs in CSHINE also carry information on the reaction dynamics. 
Thus, the CSHINE detection system is a useful tool for research on the reaction dynamics and thermodynamics of nuclear matter produced in heavy ion collisions in the Fermi energy regime.

\section{Summary}\label{sec.V}

In summary, phase I of the CSHINE detection system, which consists of two SSD telescopes, three PPACs, and three Au(Si) telescopes, was mounted and operated in the 30 MeV/u $^{40}$Ar + $^{197}$Au reaction. The data analysis focusing on the SSD telescopes demonstrated that hydrogen and helium isotopes were clearly identified using the $\Delta E-E$ method. The interstrip signal sharing in the SSSD is below $1\%$. The energy resolutions of the DSSD and CsI(Tl) are found to be approximately $1\%$ and $2\%$, respectively. The SSD telescopes were used to construct the HBT correlation functions for light charged particles. The performance was evaluated using the $\alpha-\alpha$ correlation function, and the three resonant states of $^8$Be and $^9$Be were correctly identified. CSHINE can be expected to offer opportunities for experimental studies of the collision dynamics and nuclear equation of state in heavy ion reactions at Fermi energies.

\section*{Acknowledgments}

We acknowledge the crystal group from IMP, CAS for providing the CsI(Tl) detectors, the RIBLL group for offering local help with the experiment, and the machine staff for delivering the argon beam. 

\section*{Authors' Contributions}
All authors contributed to the study conception and design. Material preparation, data collection and analysis were performed by Yi-Jie Wang, Fen-Hai Guan and Xin-Yue Diao. The first draft of the manuscript was written by Yi-Jie Wang and all authors commented on previous versions of the manuscript. All authors read and approved the final manuscript.

\end{document}